\documentclass[12pt,preprint]{aastex}

\shorttitle{Are ``EIT Waves'' Fast-Mode MHD Waves?}
\shortauthors{Wills-Davey et al.}

\begin{document}

\title{Are ``EIT Waves'' Fast-Mode MHD Waves?}

\author{M. J. Wills-Davey, C. E. DeForest, and J. O.
Stenflo\altaffilmark{1}}
\affil{Department of Space Studies, Southwest Research Institute,
        Boulder, CO 80302}

\altaffiltext{1}{On leave from the Institute of Astronomy, ETH
Zurich}

\begin{abstract}
We examine the nature of large-scale, coronal,
propagating wave fronts (``EIT waves'') and find they are incongruous with
solutions using fast-mode MHD plane-wave theory.  Specifically, we consider
the following properties: non-dispersive single pulse manifestions, observed
velocities below the local Alfv\'en speed,
and different pulses which travel at any number of constant velocities,
rather than at the ``predicted'' fast-mode speed.  We 
discuss the possibility of a soliton-like explanation for these phenomena,
and show how it is consistent with the above-mentioned aspects.   
\end{abstract}


\clearpage

\section{Introduction}
\label{sec:introduction}

Long before the availability of direct observations
in 1997 \citep{Thompson1998}, attempts were
made to explain the physics of large-scale coronal pulse waves.
The original evidence of these wave fronts appeared in chromospheric
hydrogen-$\alpha$ observations of ``Moreton waves''---semi-circular
propagating depressions, which traveled away from flaring regions at
speeds orders of magnitude above the
chromospheric sound speed \citep{Athay_Moreton1961}.  \citet{Uchida1968}
theorized that Moreton waves were a secondary effect caused by the
``skirt'' of a coronal fast-mode magnetoacoustic
shock wave extending down into the chromosphere.  They manifest themselves
in running difference images as dark fronts followed by light fronts, as shown
in Figure~\ref{fig:moreton}.

The advent of continuous soft x-ray and EUV observation---instruments 
such as {\it Yohkoh}-Soft X-ray Telescope (SXT) and the {\it SOHO} Extreme
Ultraviolet Imaging Telescope (EIT)---made it possible to test 
the \citet{Uchida1968}
theory, and indeed all manner of large-scale
coronal pulse waves have been observed.  Wave fronts have been recorded
in soft x-ray \citep{Hudson2003,Warmuth2005,Khan_Aurass2002},
EUV \citep{Thompson1998,WD_Thompson,Biesecker}, 
and even as a secondary response in
He~10830\AA\ \citep{Gilbert_etal}.  
Moreton waves have some
cospatiality with EUV waves \citep{Eto,Okamoto_etal} 
and have been particularly
well-correlated with soft x-ray observations 
\citep{Narukage2002,Narukage2004}, 
lending credence to
the original \citet{Uchida1968} postulation.  

However, Moreton waves are observed in conjunction with only a tiny
fraction of coronal observations.  The large majority
of single-pulse wave fronts
are seen only by EUV instruments, with no apparent chromospheric or
soft x-ray counterpart.  These EUV fronts (often called ``EIT waves'')
have some of the same general characteristics as Moreton
waves, but in many respects they are quite different.  While it would 
appear that Moreton waves may fit the \citet{Uchida1968} fast-mode
MHD shock model, we postulate that existing MHD models of EIT waves
are not consistent with aspects of available data, and suggest that mechanisms
which encompass nonlinear wave pulse propagation appear more promising to
explain the breadth of observed EIT wave phenomena.

\subsection{Properties of Moreton and EIT Waves}
\label{sec:properties}

Moreton waves and EIT waves can be described as ``single-pulse'' phenomena.
Figure~\ref{fig:waves} shows examples of two different EIT wave events---
one observed by {\it SOHO}-EIT, and one by the Transition Region and
Coronal Explorer ({\it TRACE}).
These waves are associated with impulsive events, 
and although actual causality has still not been determined,
\citet{Biesecker} and \citet{Cliver2005} find a strong correlation
with CME initiation.
There is also evidence that both EIT waves and Moreton waves displace
large magnetic structures: EIT waves have been observed
directly instigating loop ocillations
\citep{WD_Thompson},
and Moreton waves have been
associated with ``winking filaments''
\citep{Okamoto_etal}.

However, other aspects of EIT and Moreton waves are sufficiently different
that some have theorized they are two entirely different populations, which
originate from different instigators \citep{Chen_etal,Eto,Chen_etal2005}. 
Although both are single pulses, Moreton waves are strongly-defined,
narrow, semi-circular fronts, while EIT waves are broad ($\sim 100$~Mm),
extremely diffuse, and (when unimpeded) produce circular wave fronts.  
Moreton waves
have relatively short lifetimes (usually $< 10$ minutes), and have shown
cospatial observational signatures between the chromosphere and the
soft x-ray corona \citep{Khan_Aurass2002,Narukage2002}.
EIT waves are primarily visible in the lower corona (at 1-2~MK),
but typically have lifetimes of over an hour and can travel the entire diameter
of the Sun while remaining coherent \citep{Myers}.
Moreton waves typically travel at speeds of $\sim 400$-2000~km/s
\citep{Becker,Smith_Harvey1971}; such
velocities are thought to be comparable to or much larger than the local
Alfv\'en speed.  In recent work,
\citet{Narukage2004}---having 
calculated local fast magnetoacoustic speeds
of 700-1000~km/s---find that Moreton waves occur at speeds of
$M > 1$, and disappear as they slow to $M=1$.
EIT waves, on the other hand, travel much more slowly, at 
average velocities ranging
25-450~km/s \citep{Myers}, which correspond to $0.03 < M < 0.53$.  
Although there is some evidence of Moreton and EIT waves
traveling cospatially \citep{Thompson2000,Okamoto_etal}, most
studies conclude that, while they appear to originate together,
the two must be inherently
different \citep{Chen_etal,Eto,Chen_etal2005}. 

The work of \citet{Uchida1968} and \citet{Narukage2004} 
would appear to 
explain the nature of Moreton waves---they exist as a result of coronal
shock fronts.  EIT waves, however, have proved much more difficult to
comprehend.  Though Moreton waves are always viewed in conjunction with 
EIT waves, the converse is not true, even in high-cadence data.  
\nocite{Wills-Davey2006, Wills-Davey2002} Wills-Davey(2002,2006) 
present quantitative analysis of a {\it TRACE}-observed
EIT wave from its inception, and no corresponding Moreton wave is
observed. \footnote{This contradicts the findings of  
\citet{Harra_Sterling2003}, 
but their conclusions about the same
wave front are the result of visual inspection,
whereas the work of Wills-Davey(2002,2006) is quantitative.}

Any complete theory of EIT waves must explain:
\begin{itemize}
\item why EIT waves are observed as single pulses
\item how most EIT waves are manifested in the absence of
      Moreton waves,
\item why many EIT wave velocities are slower than predicted
      Alfv\'en speeds, 
\item why individual EIT waves travel at approximately constant speed,
      but that speed varies greatly between EIT waves, and
\item how EIT waves can maintain coherence over distances comparable 
      to the solar diameter;
\item additionally, it should confirm why EIT waves sometimes 
      generate loop oscillations.
\end{itemize}

\subsection{Existing Coronal Pulse Wave Models}
\label{sec:existing_models}

At present, multiple published explanations exist
to explain EIT waves
\citep{Chen_etal,Warmuth,Chen_etal2005,Wang,Wu_etal,
Ofman_Thompson2002,Ofman2007}.
In each case, some of the requirements listed in \S~\ref{sec:properties}
are fulfilled, but no one numerical or
theoretical model explains all six properties.

\citet{Chen_etal}, \citet{Warmuth}, 
and \citet{Chen_etal2005}
each develop models that focus on the {\it relation} between
the Moreton and EIT waves, leading to explanations 
where the Moreton wave is the primary source of a
secondary EIT wave---a scenario which is inconsistent with the bulk of
``EIT wave only'' observations. Additionally,
the models created by \citet{Chen_etal} and \citet{Chen_etal2005}
demonstrate Moreton wave propagation over large distances but EIT wave
propagation over much smaller distances---the opposite of what is seen
in observations.

In cases where ``EIT wave only'' numerical simulations have been developed,
the physics driving EIT waves is derived from the original 
\citet{Uchida1968} theory:
EIT waves are treated as fast-mode magnetohydrodynamic (MHD) waves.
Starting from this premise, 
\citet{Wang}, \citet{Wu_etal}, \citet{Ofman_Thompson2002}, and
\citet{Ofman2007} have created 
computer-generated coronal waves that imitate data very closely.
Unfortunately, the successful implementation of these models requires
unusually low quiet sun magnetic field strengths as well as a 
high plasma-$\beta$
corona.  Additionally, both the \citet{Wang} and the 
\citet{Wu_etal}
simulations reproduce only the same oft-studied event from May 1997. 
The wide variety of EIT wave velocities and morphologies
may be beyond the capability of these models; indeed, when \citet{Wang}
models a second event from April 1997, he reproduces the velocities of the
May 1997 rather than the April 1997 wave.

The problem may lie in the treatment of EIT waves as fast-mode
MHD pulses.  While a fast-mode MHD wave does have some of the properties
associated with EIT waves, many aspects of these coronal pulse waves
contradict predicted fast-mode behavior.  Additionally, 
\citet{Wills-DaveyPhD} and \citet{Warmuth2004II} 
have found observational evidence that
these waves are highly non-linear, 
with density perturbations of 40\% to more than
100\% above the local background.

In this paper, we discuss the discrepancies
between the predicted behavior of MHD waves and EUV observations,
and consider the ramifications of the MHD solution on 
other aspects of coronal physics (\S~\ref{sec:plane_wave}).  With these
discrepancies in mind, we show that aspects of
a single-pulse solution can account for the properties of 
EIT waves (\S~\ref{sec:resolution}).  

\section{Inconsistencies Arising from a Fast-Mode MHD Solution}
\label{sec:plane_wave}

At first glance, the choice of a fast-mode MHD solution seems the
most appropriate to explain EIT waves.  
Fast- and slow-mode MHD wave mode speeds can be written as
\begin{equation}
\label{eqn:fm_sm}
{v_{f,s}}^2 = \frac{1}{2}[ {v_A}^2 + {c_s}^2 \pm 
\sqrt{{v_A}^4 + {c_s}^4 - 2{c_s}^2{v_A}^2 \cos{2\theta}}]
\end{equation} 
where
${v_A}^2 = {B^2}/(4{\pi}{\rho})$ defines the local Alfv\'{e}n speed, and
${c_s}^2 = ({\gamma}{k_B}T)/m$ the local sound speed.  Note that 
$v_f \ge v_A$ and $0 \le v_s \le c_s$, depending on $\theta$.  This means 
that any event with a speed below $v_A$ cannot be considered a fast
magnetosonic wave.  

To reproduce EUV observations, the chosen wave solution
must be a compressive MHD wave that can travel
ubiquitously through a magnetized plasma.  
Pure Alfv\'en waves
cannot produce the necessary compression to be seen as a brightness
enhancement.  Slow-mode magnetoacoustic waves
are compressive, but their propagation is limited by magnetic field 
direction; the slow-mode velocity vanishes for propagation perpendicular to
field lines.  Not only would this prevent the observed ubiquitous
propagation through the quiet corona, but 
{\it TRACE} observations show evidence of 
EIT waves successfully crossing coronal loop structures \citep{WD_Thompson}.

However, fast-mode MHD waves have the double advantage of being compressive
and existing for all magnetic field orientations.  These are the properties
that led \citet{Uchida1968} to use fast-mode MHD shocks for 
his original Moreton wave solution, and motivate their continued use
in more recent simulations.
Unfortunately, a fast-mode solution presents problems when trying
to recreate some of the properties of EIT waves.  In particular,
it becomes difficult to explain:
\begin{itemize}
\item observed speeds,
\item theoretical assumptions of a low-$\beta$ corona (due the fact that
many EIT waves travel slower than the local sound speed),
\item the variety of observed propagation speeds, and
\item the nature and duration of pulse coherence.
\end{itemize}

\subsection{Velocity Magnitudes}

One inconsistency between a fast-mode 
MHD wave model and observed EIT wave behavior
concerns EIT wave speed magnitudes.   
Magnetic field orientation constrains a fast-mode MHD wave to a velocity
range $v_A \le v_{fm} \le ({v_A}^2 + {c_s}^2)^{\frac{1}{2}}$.
Previous studies of EIT and Moreton waves have defined initial conditions
such that the resultant local Alfv\'en or fast-mode
speed is also the EIT wave speed as observed for a particular event
in the data \citep{Wang,Wu_etal};
therefore, since EIT waves have been observed
at any number of speeds \citep{Myers}, we
take myriad existing work into account and
determine as large a range of quiet sun fast-mode speeds as we can from the 
data. 

Various studies have determined plasma conditions for the 
base of the quiet corona.  Magnetic field strengths 
have been measured at anywhere from 2.2~G \citep{Fludra_etal,Falconer_Davila} 
to 10~G \citep{Pauluhn_Solanki}, while multiple studies have
found density measurements close to $2 \times 10^8~\rm{cm}^{-3}$
\citep{Aschwanden_Acton,Feldman_etal,Doschek_etal}.

Such findings lead to a wide range of possible Alfv\'{e}n speeds.
\citet{Gopalswamy2002} assume a magnetic field 
strength of 2.2~G and a density of $5 \times 10^8~\rm{cm}^{-3}$,
resulting in $v_A = 215$~km/s and 
$v_{fm} = 230$~km/s at the base of the quiet corona, where $v_{fm}$ is the
fast-mode speed perpendicular to the magnetic field.  We consider
the \citet{Gopalswamy2002} velocities
a lower bound for fast-mode speeds.  By taking
the highest measured field strength (10~G) and the 
most predominantly measured 
density ($2 \times 10^8~\rm{cm}^{-3}$), we find that the Alfv\'{e}n
speed in the quiet Sun can reasonably extend as high as 1500~km/s.  
These values provide us with a (rather broad) range of possible
quiet sun Alfv\'en speeds.

Since the minimum fast-mode speed is constrained by the Alfv\'en speed,
any EIT wave must travel faster than $v_A$ for a fast-mode MHD solution to
be valid. Until now, most studies have considered sample sets weighted
towards faster waves (Gopalswamy \& Kaiser(2002); Warmuth et al.(2004a);
Narukage et al.(2005); etc.) 
\nocite{Gopalswamy2002,Warmuth2004I,Narukage2005} 
because they have focused on events correlated
with shocked Moreton waves; in such samples 
(with mean velocities of $\sim 200-400$~km/s) problems with the
fast-mode velocity are not as readily apparent.

For comparison, Figure~\ref{fig:MT_plot} shows all the mean velocities
recorded by \citet{Myers}; of the 175 EIT waves
occurring between 25 March 1997 and 16 June 1998, 160 were observed in 
multiple frames.   
Only a small fraction of the \citet{Myers} EIT waves have
average speeds close to 300~km/s; the large majority are noticeably slower.
The velocities shown in Figure~\ref{fig:MT_plot} are inconsistent with
a minimum Alfv\'en speed of 215~km/s.
101 of the 160 observed events have average speeds below
our minimum $v_A$.  Such a large discrepancy suggests that some
physical assumption is incorrect.  

A lower $v_A$ would be possible if
we found the measured values for $\mathbf{B}$ were too high and/or
the measured densities too low.  
The \citet{Lin_Kuhn} direct measurements of coronal magnetic field find 
4~G 75~Mm above an active region; presumably field strengths are lower
in the quiet corona through which the waves propagate.  
However, since the density falls off as 
$\sim e^{(-z / \Lambda)}$, where $\Lambda$ is the pressure scale height,
we actually expect the Alfv\'en speed to increase with altitude.
Using extrapolation methods, recent studies have also
found ``true'' quiet sun magnetic fields in the range of
20-40~G \citep{Krivova_Solanki,Cerdena_etal}; these values
would increase calculated Alfv\'en speeds
by as much as an order of magnitude. 
Alternatively, the problem could lie with
the assumption of a fast-mode solution.  

It may be possible that Figure~\ref{fig:MT_plot} actually shows a superposition
of two different types of wave events; there is a slight visual break at around 
$\sim 260$~km/s, suggesting we may be observing clustering of two populations.
If this is the case, it is possible the higher speed events (26 of 160) may
be consistent with fast magnetosonic wave simulations (see, for example,
Wang (2000); Wu et al. (2001); Ofman \& Thompson (2002); Ofman (2007)).   
However, the slower events would still need a separate explanation. 
\nocite{Ofman2007}

\subsection{Requirements of a Coronal Plasma}

To further emphasize the potential problems with treating EIT waves as
fast-mode waves, we consider the requirement that $v_{fm} \ge v_A$.
By assuming that EIT waves (as fast-mode waves) travel at
$v_{fm}$, we set an upper bound
for $v_A$.   According to Figure~\ref{fig:MT_plot}, this would
give us Alfv\'en speeds ranging 
$27~{\rm km/s} \le v_A \le 438~{\rm km/s}$.

We can determine the validity of these possible Alfv\'en speeds
by considering them in the context of plasma-$\beta$.  $\beta$
describes the ratio of gas pressure to magnetic pressure, and is 
often written $\beta = (8{\pi}p)/B^2$.  This also means that  
\begin{equation}
\beta \sim \frac{{c_s}^2}{{v_A}^2}.
\end{equation}
with a difference of a factor $2/{\gamma}$ (where
$\gamma$ is the ratio of specific heats), which is of order unity.
Most EIT waves are observed in the 195~\AA\ passband, which 
is most sensitive to plasma at
$\sim 1.5$~MK.  We take
the sound speed at this temperature (185~km/s) as a
reasonable value for $c_s$. 
Using the average velocities shown in Figure~\ref{fig:MT_plot}
to define $v_A$, we find that the plasma-$\beta$ 
associated with EIT waves can extend from
$\beta \sim 0.20$ to as much as $\beta \sim 50$ using these values of
$c_s$ and $v_A$.  

It has become widely accepted that coronal morphology is magnetically
dominated, and is often approximated by a force-free field.  
By definition, $\beta$ must be small in a
magnetically dominated plasma, and only for $\beta \ll 1$
is a force-free field model reasonable; 
Some recent work has discussed the possibility of coronal 
plasma-$\beta$s close to unity \citep{Gary,Aschwanden1999}, but
these measurements have typically been taken above active regions
and are assumed to apply to current-filled loops.  In any case,
it is rare to find a theoretical $\beta$ much larger than unity.

Some fully-three-dimensional Quiet Sun models 
(such as that of \citet{Wu_etal}) have implied $\beta$ can be as high as
$5 \ge \beta \ge 50$ in active region latitudes, over +/-$30^\circ$ 
(see Figure 2 of 
\citet{Wu_etal}).  While it is true that we have no direct measurements for
values of density or magnetic field in the corona, under such conditions (and
over such an extended area), the morphology of the corona in EUV images
demonstrates that the $\beta$ parameter must be low; magnetic structures
dominate everywhere.

Until observations imply that there are extensive (of order several hundred Mm)
areas of the corona with such high $\beta$, we will instead be swayed by 
existing coronal limb observations.
If the $\beta$s found using the \citet{Myers}
data are valid, then large
portions of the quiet sun cannot be magnetically-dominated.
Such large possible $\beta$ values either contradict 
the validity of the corona as 
a low-$\beta$ plasma, or offer additional evidence that EIT waves 
cannot be modeled using fast-mode waves.

\subsection{Propagation Speed Differences}

The fact that a broad range of speeds is observed at all should 
cause us to question the validity of fast-mode waves 
as an EIT wave solution.  In a linear 
regime, the wave speed corresponds to the reaction speed of the medium;
wave velocities are directly correlated to observable properties such as
density or magnetic field strength.  Observations of EIT waves show that 
pulses maintain coherence over global distances. 
This lack of decoherence suggests that 
the plasma properties of the quiet corona 
are often uniform.  If EIT waves were actually fast-mode MHD waves,
this underlying global sameness would constrain EIT waves to a narrow
range of velocities close to the expected fast-mode speed.  The
simulations of \citet{Wang}  
and \citet{Wu_etal}, which propagate
fast-mode waves through mildly-structured quiet corona, produce just
this type of result; \citet{Wang}
finds that even quiet sun changes
over time are not large enough to substantially affect the fast-mode speed.

While EIT observations lack the temporal cadence to show if EIT waves
travel at constant speed, the range of speeds found by 
\citet{Myers}
makes it difficult to justify the existence of a
``preferred'' EIT wave speed.  Additionally, the \citet{Myers} data
show strong evidence of
waves with different speeds traveling through the same region of quiet
sun in the space
of a few hours.  In the case of one particularly productive active region,
seven waves were produced over a 36-hour period (from 1-May to 3-May 1998) 
with speeds of 85 to 435~km/s, a difference of a factor
of 5.  In each case, the wave traveled a distance of $\sim 1~R_{\odot}$
through the same general area of quiet Sun.  

Explaining each of these wave fronts as fast-modes would require
that the quiet sun fast-mode speed change globally on time-scales
shorter than a few hours.  Since EIT waves are strongly associated with
CMEs, it may be that CMEs corresponding to EIT waves
produce large-scale topology changes which then affect the global 
fast-mode speed.  However, the lack of global changes shown by difference
images suggests that this is unlikely.

\subsection{Pulse Coherence}
\label{sec:pulse}

Morphologically, EIT waves appear as single-pulse fronts.  
To date, there has only been one observation of a pulse wave
(in this case, a Moreton wave) that appears to include multiple fronts, 
related to the X10 flare of 29 October 2003 \citep{Neidig2004}; 
unfortunately, no contemporaneous EUV data exist.  

Numerical simulations have shown that a fast-mode MHD solution can 
generate a wave packet comparable to EUV observations.  
A wave packet of fast-mode MHD waves can produce a single-pulse front; 
however, the differing phase speeds within the packet
would leave it highly susceptible to
dispersion resulting from conditions such as density stratification 
and magnetic field variations.  If the scale of the fluctuations is 
much smaller than the wavelength (as in the case of magnetic 
field loops), these fluctuations will not affect the coherence of the 
wave; however, the pulse width appears to be about a scale height
\citep{Wills-DaveyPhD}, allowing for significant effects due to density
variations.  Since the dispersing medium is ubiquitous, 
the packet would begin to break apart almost immediately, and 
would appear as periodic ``ripples'' on either side of the main front.

Such immediate dispersion effects appear difficult to reconcile with 
observations of single coherent fronts propagating over global distances.
The lack of temporal resolution in the {\it SOHO}-EIT data may account for
the lack of any observed periodicity (perhaps visible as multiple fronts)
as the front widens and the amplitude decreases.  However, 
multiple {\it TRACE}
observations, despite a much higher cadence, have also failed to
reveal any obvious periodicity in a wave as it decays.

Quantitative measurements of the 13 June 1998 EIT wave 
(Figure~\ref{fig:waves}(b)) show that the
density enhancement cross-section maintains coherence for some
time (of the order of tens of minutes), and will even break apart slightly 
and re-form in a ``pulse'' shape as it encounters different coronal structures 
\citep{Wills-Davey2006}.   
In this quantifiable case, the wave amplitude decreases over 
time---in a manner consistent with radial expansion---but 
there is no measurable increase in the pulse FWHM \citep{Wills-DaveyPhD}. 
Additionally, no ``ripples''
appear around the main pulse as this occurs.  
To the extent that the pulse was measurable
(before it became indistinguishable from noise),
the data appear to be consistent with a wave
propagating dispersionlessly.

This lack of dispersion
also appears consistent with the wavelet analysis performed by
\citet{Ballai2005}.  
Over the length of the entire 13 June 1998 data set,
their results show a roughly constant wavelet power spectrum band 
ranging from 285 to 560 seconds.  They interpret this as a strong signal
with an intensity period $\sim 400$~seconds that {\it does not degrade}.  
While their results do not shed light on 
the nature of the pulse-like structure
of the front, they do appear to confirm that the wave packet remains
intact throughout their measurements.  Given some of the interference
seen in the \citet{Wills-Davey2006} cross-sections, 
the \citet{Ballai2005} findings would suggest that
the pulse is unusually stable to perturbations, and does not suffer from
the dispersion expected for a linear fast-mode wave.

\section{Resolving EIT Wave Inconsistencies}
\label{sec:resolution}

EIT wave velocities have presented two key problems: the speeds are too
slow for a significant number of observations
to be explained using fast-mode MHD waves, and the plasma properties
of a largely uniform quiet corona should not lead to such a wide range
of constant wave speeds.  However, if we instead understand EIT waves as
a type of coronal MHD soliton---perhaps a 2-D slow-mode soliton---the 
velocity range becomes easier to explain.  

One key difference between plane wave and
soliton solutions is the velocity dependence.  With a linear MHD solution,
wave speed is determined solely by properties of the transmission medium.
Soliton speed is additionally dependent on the amplitude of the
pulse.  In the case of MHD solitons,
speed varies as a function of density enhancement
\citep{Buti,Ballai2003}.

Consider the velocity dependence shown in
Figure~\ref{fig:MT_plot}.  Although the ``Quality Rating'' is a 
visually-determined observer-dependent ranking system, the data from
\citet{Myers} still show that well-defined
(more density-enhanced) waves to travel faster. 
Speeds only approach, but do not reach, 
the Alfv\'en speed $v_A$; \citet{Narukage2004}
show that large-scale pulse waves traveling at or above $v_A$ 
shock and appear as Moreton waves.  The velocity-density enhancement
dependence also allows for events of different speeds to pass
through the same region of quiet sun without requiring global
restructuring.

In addition to solving the velocity discrepancies, a soliton explanation
also provides some of the pulse stability and coherence needed to
explain the properties of EIT waves.  Because the stability of a soliton
is dependent both on both nonlinearity in the pulse and dispersion in the
local medium, solitons are stable to small perturbations, allowing
them to travel through thin cross-wise loop structures 
and over large distances of quiet sun.
As solitons are nondispersive, this explanation would
also consider the lack of dispersion observed in 
strong events such as the 13 June 1998 event (Figure~\ref{fig:waves}(b)).

While it is true that the slow-mode experiences greater dissipation than 
other modes, because of their large width, EIT waves only need to remain 
coherent over $\sim 10$ wavelengths to display typical behavior.  If we 
consider the work of \citet{Ofman1999},
who looked at slow MHD waves
in polar plumes, they found that pulses maintained coherence for a minimum of 
three wavelengths, and showed the sort of steepening that would be
counteracted by nonlinearity, in the case of soliton behavior.  This suggests
that coherence over ten wavelengths is not unreasonable.

Of course, the dynamics observed in EIT waves 
could not be the same as those seen by
\citet{Ofman1999}.  Rather, since these wave fronts propagate 
laterally through the corona and are
at least a scale height tall, they will rely on a different 
steepening/dispersion mechanism to create soliton-like behavior.
The steepening could come from the fact that Alfv\'en speed increases with
altitude in the corona.  The
dispersion mechanism could be lateral density stratification across the pulse
itself.  The fact that the medium is itself MHD also means that the wave must
have a magnetic component.  However, since such a large pulse must 
remain coherent to small perturbations (such as magnetic loops), it may imply
that only very strong (i.e. active regions) or very defined (i.e. coronal
holes) magnetic structures have any noticeable effect on the wave.

The effect of a soliton on the local medium can also account for
loop oscillations.  As a compressive wave packet with no related rarefaction,
it must displace the medium 
in the direction of propagation, where the displacment will remain
unless restored
by some other force.  While this argument can account for any linear
compressive wave packet, it is still consistent with the effects
of a soliton.
In the case of an EIT wave, coronal material will be carried with the front.
\citet{WD_Thompson} 
demonstrated this for the 13 June 
1998 event, as they tracked individual loops along with the wave.  However,
since the magnetic fields of the corona are anchored in the photosphere,
after the wave has moved on, the individual loops will ``snap'' back.
\citet{Wills-DaveyPhD}
found that most structures behaved in an overdamped
manner when returning to their original positions, but some 
loops---often aligned perpendicular to the direction of propagation---showed
oscillatory behavior.  

Lastly, the production of a soliton does not require the presence of a
shock, allowing for the existance of EIT waves in the absence of 
Moreton waves.  While this still doesn't explain the relationship 
between Moreton and EIT waves, a soliton-like EIT wave can
account for the vast majority of observations.

\section{Discussion}
\label{sec:discussion}

The consistency of the properties of EIT waves has long
motivated solar physicists to develop a physical understanding as to
their nature.  Developing this understanding has proved elusive
in previous work.  Unfortunately, fast MHD compressional waves do not properly
describe dynamics of many EIT wave events.
The physical properties of EIT waves---their single-pulse, stable
morphology; the non-linearity of their density perturbations; the
lack of a single representative velocity---instead suggest that they may be
best explained as soliton-like phenomena.  

While most fronts travel below the expected coronal Alfv\'en speed, 
as a general trend, larger density perturbations tend to move at
faster velocities.  
There is also the observational evidence that many EIT wave
pulse widths are close to one to two scale heights; 
this may be a visual effect, but is is possible that pressure and magnetic
forces convolve to act as a wave guide, as predicted by
\citet{Nye_Thomas}.  It would be consistent with
initial findings that flux is conserved as
EIT waves propagate radially along the
solar surface rather than spherically
\citep{Wills-DaveyPhD}.  It might also account, at least in part, 
for the strong discrepancy between the number of 
EIT and SXT pulse wave observations
\citep{Sterling_Hudson,Biesecker,Warmuth2005};
SXT preferentially observes hotter structures with larger scale heights,
and the maximum pulse height of EIT waves might be constrained by
smaller, cooler loops.

The flux conservation found by \citet{Wills-DaveyPhD}
does demonstrate one
surprising, unsoliton-like behavior: after a pulse stops forming, its 
amplitude appears to drop off as $r^{-1}$.
This occurs in spite of the
other, soliton-like properties observed in EIT waves.  
It is likely that flux conservation is a necessary aspect of 
radially-propagating trapped MHD solitons, and that a decrease in amplitude
is necessary to a conservative solution.   

We feel the solitary wave hypothesis 
offers the most compelling explanation to date
for the properties of EIT waves.  While the derivation of a two-dimensional
MHD soliton solution is perhaps beyond analytical scope, and therefore must
be demonstrated numerically, the properties inherent in a soliton-like 
explanation should
fit the data much better than the oft-used fast-mode solutions.
It becomes possible to explain:
the lack of a ``typical'' EIT wave velocity;
the amplitude-velocity relationship seen by \citet{Myers};
and the consistent observations of single, coherent, nonlinear coronal pulses.
While we do not pretend to offer a comprehensive explanation on the nature of
EIT waves, by offering this interpretation, we hope to assist theorists and
modelers by providing a new direction.

Developing a consistent theoretical understanding of EIT waves is particularly
important in the context of new EUV missions.  {\it TRACE}
observations have already shown that wave parameters are easily quantitatively 
measured with sufficient spatiotemporal resolution \citep{Wills-Davey2006}.  
If we can correctly
model and reproduce EIT waves,  we can use wave properties
extracted from observations to inverse model the plasma parameters of the
affected quiet corona.  Missions like {\it STEREO} \citep{STEREO} and 
{\it GOES-N} \citep{GOES-N} will
give us the opportunity to develop and test these observational modeling
tools.  

Using large-scale propagating waves for 
{\it global coronal seismology} has been postulated since \citet{Meyer}.
However, its successful implementation requires a well-understood 
theoretical model.  Previous studies \citep{Meyer,Ballai2003}
have attempted to calculate quiet sun magnetic field strengths with
global coronal seismology using the assumption that Moreton waves and
EIT waves can be modeled as MHD fast-mode waves.  \citet{Meyer} finds
a field strength that appears high; \citet{Ballai2003}
finds one that
appears low.  An accurate wave model 
may result in a more reasonable field strength calculation, allowing
EIT waves to make the transition
from coronal phenomena to observational tool.

\begin{figure}
\plotone{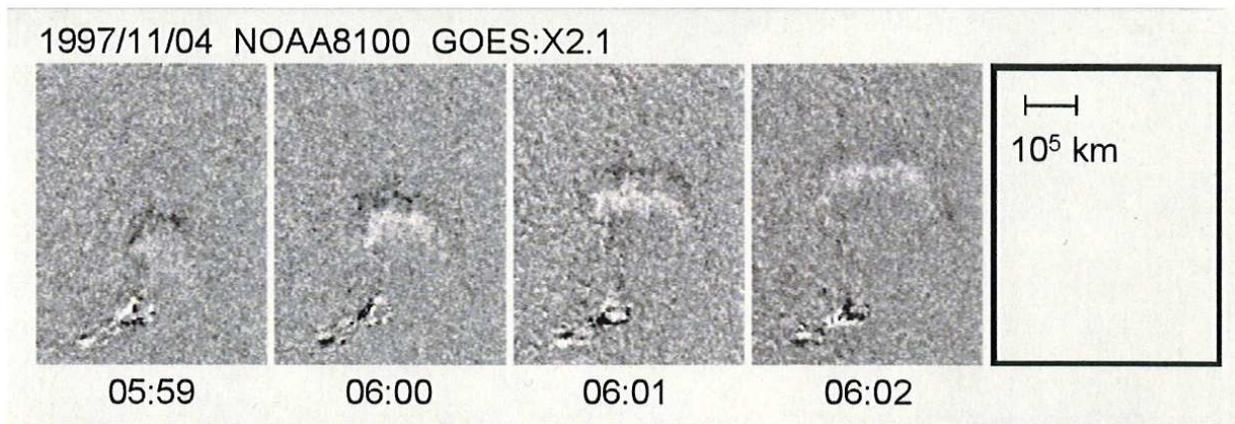}
\caption{An example of a Moreton wave observed in H-$\alpha$
on 4 November 1997 by the Flare Monitoring
Telescope of Kyoto University's Hida Observatory.  This event was produced
in conjunction with a GOES~X2.1 flare.
{\it This figure reproduced from \citet{Narukage2003}
courtesy of N. Narukage.}}
\label{fig:moreton}
\end{figure}

\begin{figure}
\plotone{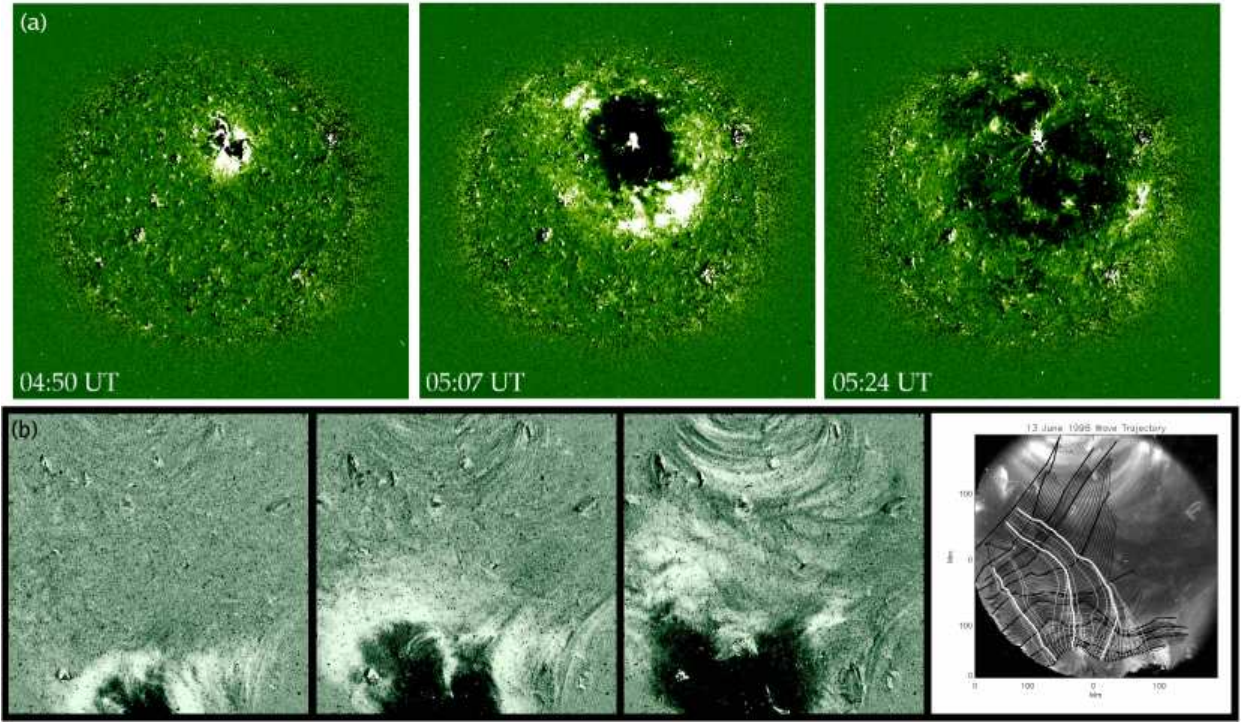}
\caption{Two examples of EIT waves as observed by different EUV
instruments.  (a) shows running difference images of
an EIT wave seen by {\it SOHO}-EIT on 12 May 1997,
and studied in detail by
\citet{Thompson1998}.  (b) shows base difference images
and measured fronts from \citet{Wills-Davey2006} of an event observed
by {\it TRACE} on 13 June 1998.
(In running
difference images, each frame is subtracted from the one following.
In base difference images, all frames have a single pre-event image
subtracted from them.)
{\it Figure~\ref{fig:waves}(a) reproduced courtesy of B. J. Thompson.}}
\label{fig:waves}
\end{figure}

\begin{figure}
\plotone{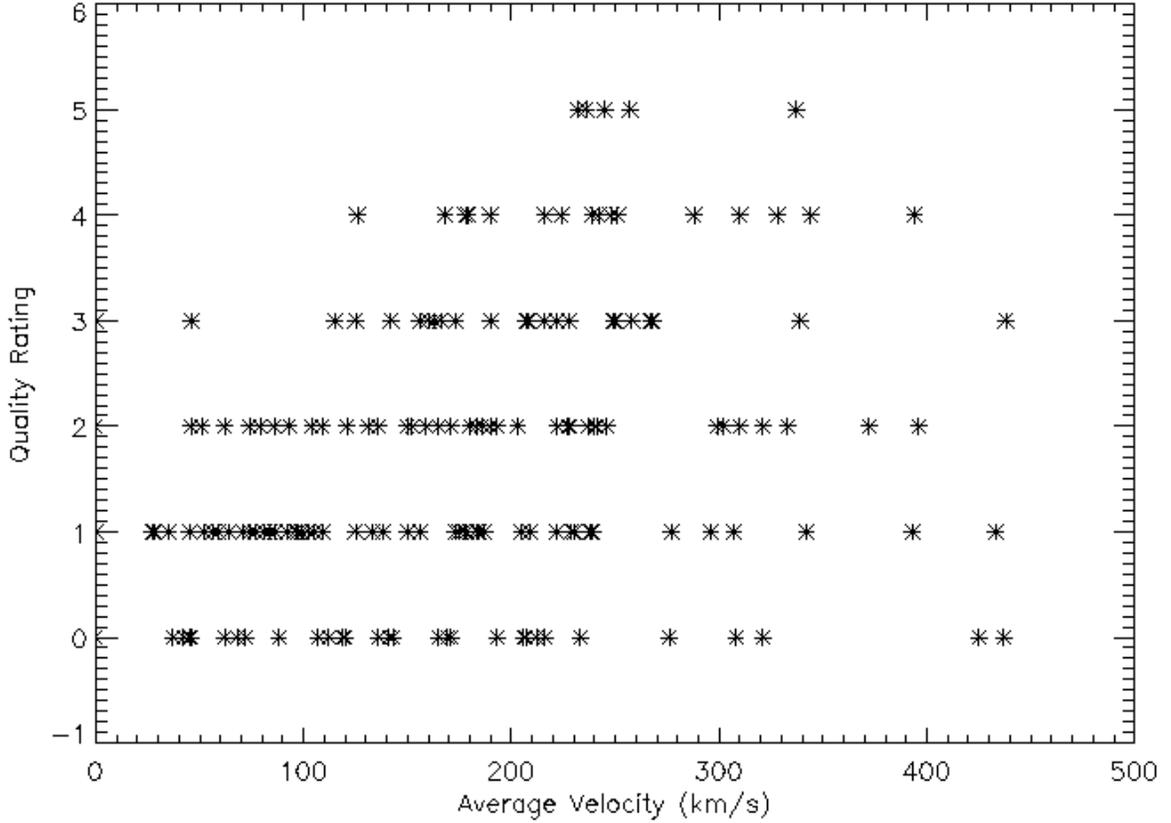}
\caption{Distribution of average EIT wave speeds with respect to
``Quality Rating,'' a subjective measure corresponding to the observer's
confidence in the velocity reading, with zero being a low confidence
score \citep{Myers}.  While ``Quality Ratings''
have no quantifiable validity, these data suggest a correlation between
density enhancement and speed.  Note the substantial velocity spread, with
events traveling at an average speed in the range $25 < v < 438$~km/s, and
many traveling below the minimum calculated Alfv\'en speed of 215~km/s.}
\label{fig:MT_plot}
\end{figure}

\acknowledgements
The authors wish to thank V. J. Pizzo for insightful comments and editting.
Figure~\ref{fig:waves}(a) was reproduced by permission of B. J. Thompson.
Figure~\ref{fig:moreton} was reproduced by permission of N. Narukage.
This research was funded by NASA Grant LWS 02-0000-0025.

\clearpage

\bibliographystyle{apj.bst}
\bibliography{ms.bib}

\begin{thebibliography}{}

\bibitem[\protect\citeauthoryear{Aschwanden \& Acton}{Aschwanden \&
  Acton}{2001}]{Aschwanden_Acton}
Aschwanden, M.~J.,  \& Acton, L.~W. 2001, ApJ, 550, 475

\bibitem[\protect\citeauthoryear{Aschwanden et~al.}{Aschwanden
  et~al.}{1999}]{Aschwanden1999}
Aschwanden, M.~J., Newmark, J.~S., Delaboudini\`ere, J.-P., Neupert, W.~M.,
  Klimchuk, J.~A., Gary, G.~A., Portier-Fozzani, F.,  \& Zucker, A. 1999, ApJ,
  515, 842

\bibitem[\protect\citeauthoryear{Athay \& Moreton}{Athay \&
  Moreton}{1961}]{Athay_Moreton1961}
Athay, R.~G.,  \& Moreton, G.~E. 1961, ApJ, 133, 935

\bibitem[\protect\citeauthoryear{Ballai, Erd\'elyi, \& Pint\'er}{Ballai
  et~al.}{2005}]{Ballai2005}
Ballai, I., Erd\'elyi, R.,  \& Pint\'er, B. 2005, ApJ, 633, L145

\bibitem[\protect\citeauthoryear{Ballai, Thelen, \& Roberts}{Ballai
  et~al.}{2003}]{Ballai2003}
Ballai, I., Thelen, J.~C.,  \& Roberts, B. 2003, A\&A, 404, 701

\bibitem[\protect\citeauthoryear{Becker}{Becker}{1958}]{Becker}
Becker, U. 1958, Zeitschrift f{\"{u}}r Astrophysik, 44, 243

\bibitem[\protect\citeauthoryear{Biesecker et~al.}{Biesecker
  et~al.}{2002}]{Biesecker}
Biesecker, D.~A., Myers, D.~C., Thompson, B.~J., Hammer, D.~M.,  \& Vourlidas,
  A. 2002, ApJ, 569, 1009

\bibitem[\protect\citeauthoryear{Buti}{Buti}{1991}]{Buti}
Buti, B. 1991, GRL, 18, 809

\bibitem[\protect\citeauthoryear{Chen, Fang, \& Shibata}{Chen
  et~al.}{2005}]{Chen_etal2005}
Chen, P.~F., Fang, C.,  \& Shibata, K. 2005, ApJ, 622, 1202

\bibitem[\protect\citeauthoryear{Chen et~al.}{Chen et~al.}{2002}]{Chen_etal}
Chen, P.~F., Wu, S.~T., Shibata, K.,  \& Fang, C. 2002, ApJ, 572, L99

\bibitem[\protect\citeauthoryear{Cliver et~al.}{Cliver
  et~al.}{2005}]{Cliver2005}
Cliver, E.~W., Laurena, M., Storini, M.,  \& Thompson, B.~J. 2005, ApJ, 631,
  604

\bibitem[\protect\citeauthoryear{Dom\'inguez Cerde\~na, S\'anchez~Almeida, \&
  Kneer}{Dom\'inguez Cerde\~na et~al.}{2003}]{Cerdena_etal}
Dom\'inguez Cerde\~na, I., S\'anchez~Almeida, J.,  \& Kneer, F. 2003, A\&A,
  407, 741

\bibitem[\protect\citeauthoryear{Doscheck et~al.}{Doscheck
  et~al.}{1997}]{Doschek_etal}
Doscheck, G.~A., Warren, H.~P., Laming, J.~M., Mariska, J.~T., Wilhelm, K.,
  Lemaire, P., Sch\"uhle, U.,  \& Moran, T.~G. 1997, ApJ, 482, L109

\bibitem[\protect\citeauthoryear{Eto et~al.}{Eto et~al.}{2002}]{Eto}
Eto, S., et~al. 2002, PASJ, 54, 481

\bibitem[\protect\citeauthoryear{Falconer \& Davila}{Falconer \&
  Davila}{2001}]{Falconer_Davila}
Falconer, D.~A.,  \& Davila, J.~M. 2001, ApJ, 547, 1109

\bibitem[\protect\citeauthoryear{Feldman et~al.}{Feldman
  et~al.}{1999}]{Feldman_etal}
Feldman, U., Doschek, G.~A., Sch\"uhle, U.,  \& Wilhelm, K. 1999, ApJ, 518, 500

\bibitem[\protect\citeauthoryear{Fludra et~al.}{Fludra
  et~al.}{2002}]{Fludra_etal}
Fludra, A., Ireland, J., Del~Zanna, G.,  \& Thompson, W.~T. 2002, Advanc. in
  Sp. Res., 29, 361

\bibitem[\protect\citeauthoryear{Gary}{Gary}{2001}]{Gary}
Gary, G.~A. 2001, Sol. Phys., 203, 71

\bibitem[\protect\citeauthoryear{Gilbert et~al.}{Gilbert
  et~al.}{2004}]{Gilbert_etal}
Gilbert, H.~R., Holzer, T.~E., Thompson, B.~J.,  \& Burkepile, J.~T. 2004, ApJ,
  607, 540

\bibitem[\protect\citeauthoryear{Gopalswamy \& Kaiser}{Gopalswamy \&
  Kaiser}{2002}]{Gopalswamy2002}
Gopalswamy, N.,  \& Kaiser, M.~L. 2002, Advanc. in Sp. Res., 29, 307

\bibitem[\protect\citeauthoryear{Harra \& Sterling}{Harra \&
  Sterling}{2003}]{Harra_Sterling2003}
Harra, L.~K.,  \& Sterling, A.~C. 2003, ApJ, 587, 429

\bibitem[\protect\citeauthoryear{Hudson et~al.}{Hudson
  et~al.}{2003}]{Hudson2003}
Hudson, H.~S., Khan, J.~I., Lemen, J.~R., Nitta, N.~V.,  \& Uchida, Y. 2003,
  Sol. Phys., 212, 121

\bibitem[\protect\citeauthoryear{Kaiser}{Kaiser}{2005}]{STEREO}
Kaiser, M.~L. 2005, Advanc. in Sp. Res., 36, 1483

\bibitem[\protect\citeauthoryear{Khan \& Aurass}{Khan \&
  Aurass}{2002}]{Khan_Aurass2002}
Khan, J.~I.,  \& Aurass, H. 2002, A\&A, 383, 1018

\bibitem[\protect\citeauthoryear{Krivova \& Solanki}{Krivova \&
  Solanki}{2004}]{Krivova_Solanki}
Krivova, N.~A.,  \& Solanki, S.~K. 2004, A\&A, 417, 1125

\bibitem[\protect\citeauthoryear{Lin, Kuhn, \& Coulter}{Lin
  et~al.}{2004}]{Lin_Kuhn}
Lin, H., Kuhn, J.~R.,  \& Coulter, R. 2004, ApJ, 613, L177

\bibitem[\protect\citeauthoryear{Meyer}{Meyer}{1968}]{Meyer}
Meyer, F. 1968, in Structure and Development of Solar Active Regions, IAU, 485

\bibitem[\protect\citeauthoryear{Narukage et~al.}{Narukage
  et~al.}{2005}]{Narukage2005}
Narukage, N., Eto, S., Kadota, M., Katai, R., Kurokawa, H.,  \& Shibata, K.
  2005, IAU Symp., 223, 367

\bibitem[\protect\citeauthoryear{Narukage et~al.}{Narukage
  et~al.}{2002}]{Narukage2002}
Narukage, N., Hudson, H.~S., Morimoto, T., Akiyama, S., Kitai, R., Kurokawa,
  H.,  \& Shibata, K. 2002, ApJ, 572, L109

\bibitem[\protect\citeauthoryear{Narukage et~al.}{Narukage
  et~al.}{2004}]{Narukage2004}
Narukage, N., Morimoto, T., Kadota, M., Kitai, R., Kurokawa, H.,  \& Shibata,
  K. 2004, PASJ, 56, L5

\bibitem[\protect\citeauthoryear{Narukage et~al.}{Narukage
  et~al.}{2003}]{Narukage2003}
Narukage, N., Shibata, K., Eto, S., Morimoto, T., Kadota, M., Kitai, R.,  \&
  Kurokawa, H. 2003, in SOHO 13: Waves, Oscillations, and Small Scale Events in
  the Solar Atmosphere

\bibitem[\protect\citeauthoryear{Neidig}{Neidig}{2004}]{Neidig2004}
Neidig, D. 2004, in AAS Meeting 204, \#47.10

\bibitem[\protect\citeauthoryear{Nye \& Thomas}{Nye \&
  Thomas}{1976}]{Nye_Thomas}
Nye, A.~H.,  \& Thomas, J.~H. 1976, ApJ, 204, 573

\bibitem[\protect\citeauthoryear{Ofman}{Ofman}{2007}]{Ofman2007}
Ofman, L. 2007, ApJ, 655, 1134

\bibitem[\protect\citeauthoryear{Ofman, Nakariakov, \& DeForest}{Ofman
  et~al.}{1999}]{Ofman1999}
Ofman, L., Nakariakov, V.~M.,  \& DeForest, C.~E. 1999, ApJ, 514, 441

\bibitem[\protect\citeauthoryear{Ofman \& Thompson}{Ofman \&
  Thompson}{2002}]{Ofman_Thompson2002}
Ofman, L.,  \& Thompson, B.~J. 2002, ApJ, 574, 440

\bibitem[\protect\citeauthoryear{Okamoto et~al.}{Okamoto
  et~al.}{2004}]{Okamoto_etal}
Okamoto, T.~J., Nakai, H., Keiyama, A., Narukage, N., Satoru, U., Kitai, R.,
  Kurokawa, H.,  \& Shibata, K. 2004, ApJ, 608, 1124

\bibitem[\protect\citeauthoryear{Pauluhn \& Solanki}{Pauluhn \&
  Solanki}{2003}]{Pauluhn_Solanki}
Pauluhn, A.,  \& Solanki, S.~K. 2003, A\&A, 407, 359

\bibitem[\protect\citeauthoryear{Smith \& Harvey}{Smith \&
  Harvey}{1971}]{Smith_Harvey1971}
Smith, S.~F.,  \& Harvey, K.~L. 1971, in Physics of the Solar Corona, Vol.~27,
  156

\bibitem[\protect\citeauthoryear{Sterling \& Hudson}{Sterling \&
  Hudson}{1997}]{Sterling_Hudson}
Sterling, A.~C.,  \& Hudson, H.~S. 1997, ApJ, 491, L55

\bibitem[\protect\citeauthoryear{Thompson \& Myers}{Thompson \&
  Myers}{2006}]{Myers}
Thompson, B.~J.,  \& Myers, D.~C. 2006, ApJ Supp. Series, in press

\bibitem[\protect\citeauthoryear{Thompson et~al.}{Thompson
  et~al.}{1998}]{Thompson1998}
Thompson, B.~J., Plunkett, S.~P., Gurman, J.~B., Newmark, J.~S., St.~Cyr,
  O.~C., Michels, D.~J.,  \& Dolaboudini\`{e}re, J.-P. 1998, GRL, 25, 2461

\bibitem[\protect\citeauthoryear{Thompson et~al.}{Thompson
  et~al.}{2000}]{Thompson2000}
Thompson, B.~J., Reynolds, B., Aurass, H., Gopalswamy, N., Gurman, J.~B.,
  Hudson, H.~S., Martin, S.~F.,  \& St.~Cyr, O.~C. 2000, Sol. Phys., 193, 161

\bibitem[\protect\citeauthoryear{Uchida}{Uchida}{1968}]{Uchida1968}
Uchida, Y. 1968, Sol. Phys., 4, 30

\bibitem[\protect\citeauthoryear{Wang}{Wang}{2000}]{Wang}
Wang, Y.-M. 2000, ApJ, 543, L89

\bibitem[\protect\citeauthoryear{Warmuth, Mann, \& Aurass}{Warmuth
  et~al.}{2005}]{Warmuth2005}
Warmuth, A., Mann, G.,  \& Aurass, H. 2005, ApJ, 626, L121

\bibitem[\protect\citeauthoryear{Warmuth et~al.}{Warmuth
  et~al.}{2001}]{Warmuth}
Warmuth, A., Vr\v{s}nak, B., Aurass, H.,  \& Hanslmeier, A. 2001, ApJ, 560,
  L105

\bibitem[\protect\citeauthoryear{Warmuth et~al.}{Warmuth
  et~al.}{2004a}]{Warmuth2004I}
Warmuth, A., Vr\v{s}nak, B., Magdaleni\'c, J., Hanslmeier, A.,  \& Otruba, W.
  2004a, A\&A, 418, 1101

\bibitem[\protect\citeauthoryear{Warmuth et~al.}{Warmuth
  et~al.}{2004b}]{Warmuth2004II}
Warmuth, A., Vr\v{s}nak, B., Magdaleni\'c, J., Hanslmeier, A.,  \& Otruba, W.
  2004b, A\&A, 418, 1117

\bibitem[\protect\citeauthoryear{Wilkinson}{Wilkinson}{2005}]{GOES-N}
Wilkinson, D. 2005, in AGU Fall Meeting, \#SH11A-0240

\bibitem[\protect\citeauthoryear{Wills-Davey}{Wills-Davey}{2002}]{Wills-Davey2%
002}
Wills-Davey, M.~J. 2002, in COSPAR Colloquia Series, Vol.~13, Multi-Wavelength
  Observations of Coronal Structure and Dynamics, 299

\bibitem[\protect\citeauthoryear{Wills-Davey}{Wills-Davey}{2003}]{Wills-DaveyP%
hD}
Wills-Davey, M.~J. 2003, Ph.D. thesis, Montana State University

\bibitem[\protect\citeauthoryear{Wills-Davey}{Wills-Davey}{2006}]{Wills-Davey2%
006}
Wills-Davey, M.~J. 2006, ApJ, 645, 757

\bibitem[\protect\citeauthoryear{Wills-Davey \& Thompson}{Wills-Davey \&
  Thompson}{1999}]{WD_Thompson}
Wills-Davey, M.~J.,  \& Thompson, B.~J. 1999, Sol. Phys., 190, 467

\bibitem[\protect\citeauthoryear{Wu et~al.}{Wu et~al.}{2001}]{Wu_etal}
Wu, S.~T., Zheng, H., Wang, S., Thompson, B.~J., Plunkett, S.~P., Zhao, X.~P.,
  \& Dryer, M. 2001, JGR, 106, 25,089

\end{thebibliography}

\clearpage

\end{document}